\documentclass[iop]{emulateapj}

\include{packages}

\providecommand{\nolinkurl}[1]{\url{#1}}

\begin{document}
	
	\title{The GN-z11-Flash Event Can be a Satellite Glint}
	
	\author{Guy Nir\altaffilmark{1}}
	\author{Eran O.~Ofek\altaffilmark{1}}
	\author{Avishay Gal-Yam\altaffilmark{1}}

	\altaffiltext{1}{Benoziyo Center for Astrophysics, Weizmann Institute
		of Science, 76100 Rehovot, Israel}
	
	\begin{abstract}
		
		Recently Jiang et al.~reported the discovery of a possible 
		short duration transient, detected in a single image, 
		spatially associated with a $z\sim 11$ galaxy.
		Jiang et al.~and Kahn et al.~suggested the transient originates from a $\gamma$-Ray Burst (GRB), 
		while Padmanabhan \& Loeb argued the flash is consistent with a supernova shock breakout event
		of a 300\,$M_\sun$ population III star. 
		Jiang et al.~argued against the possibility 
		that this event originated from light reflected off a satellite. 
		Here we show that reflection of sunlight from a
		high-orbit satellite or a piece of space debris
		is a valid and reasonable explanation. 
		As shown in recent works, the rate of point-like satellite reflections,
		brighter than 11th magnitude, 
		is $>10$\,deg$^{-2}$\,day$^{-1}$ 
		near the equatorial plane. 
		At higher declinations the rate is 5--50 times lower,
		but still significant:
		about four orders of magnitudes higher than the rate estimated for GRBs. 
	\end{abstract}
	
	
	\section{introduction}
	
	Recently \cite{GRB_flash_high_redshift_galaxy_Jiang_2020} presented a short duration flash 
	in a MOSFIRE \citep{MOSFIRE_system_overview_McLean_2008} $K$-band spectrum of a galaxy at redshift $z\sim11$ 
	\citep{high_redshift_galaxy_Oesch_2016}.  
	The galaxy is referred to as GN-z11, and the transient is designated GN-z11-flash. 
	The transient specific flux was 0.057\,mJy in the $K$-band, 
	and it was visible in a single 179\,s exposure. 
	\cite{GRB_flash_high_redshift_galaxy_Jiang_2020} 
	argue the source of the transient was a GRB prompt UV flash. 
	\cite{GRB_flash_context_Kahn_2020} further showed that 
	the properties of the flash are consistent with a 
	prompt optical emission from a GRB. 
	
	Shortly after, \cite{more_probable_explanation_Steinhardt_2021} suggested a more probable explanation, 
	by presenting other MOSFIRE $K$-band images where flashes are seen, 
	arguing that the source of these transients were reflections off artificial satellites. 
	In response, \cite{refuting_more_probable_explanation_Jiang_2021}
	argue that such flashes are typical of Low Earth Orbit (LEO) satellites, 
	and that those would not be a viable explanation for GN-z11-flash, 
	as it was 
	(a) a point-like object in the raw image, and 
	(b) observed inside the Earth's shadow, as seen by satellites up to 4000\,km above the ground. 
	\cite{GRB_flash_high_redshift_galaxy_Jiang_2020} also exclude a high-orbit satellite 
	as the cause of GN-z11-flash as the observation was made at a declination of $\sim 60^\circ$. 
	
	However, both \cite{geosync_satellite_foreground_Nir_2020} and \cite{satellite_glints_EvryScope_Corbett_2020} 
	presented samples of high orbit satellite glints that can explain GN-z11-flash. 
	They showed that high orbit satellites, or space debris, 
	rotating at a period of a few minutes, 
	can cause $\mathcal{O}(0.1\text{s})$ flashes, 
	that would appear as point sources under reasonable seeing conditions. 	
	These flashes would be indistinguishable from fast astrophysical transients 
	in single images. 
	
	\section{Analysis}
	
	To appear point like in a slit image, an object moving at an angular velocity $v$
	and angle $\theta$ to the slit, under $s$ seeing size, 
	must have a flash shorter than $\Delta t \lesssim s/(v\cos{\theta})$. 
	For GN-z11-flash, we take\footnote{
		We take $s=0.6''$ as this was the seeing during the observation of the flash. 
		We take $v=15''$\,s$^{-1}$ for a geosynchronous satellite moving directly overhead. 
		The actual transverse velocity could be lower if the object is closer to the horizon
		or if its orbit is eccentric and spends more time at higher declinations, 
		(e.g., Tundra orbits). 
	}	
	$s=0.6''$, $v=15''/$s and $30<\theta<60^\circ$, 
	giving $0.05 < \Delta t < 0.08$\,s. 
	Such short glints appear as one- or two-frame flashes, 
	that were fairly common in the sample presented by \cite{geosync_satellite_foreground_Nir_2020}. 
	
	The rate given by \cite{geosync_satellite_foreground_Nir_2020} is 30--40 per day per deg$^2$, 
	or $1.2\times 10^6$--$1.6\times 10^6$ per day per sky, 
	for satellites close to the equatorial plane. 
	This includes glints of 11th magnitude or brighter. 
	A 0.1\,s duration glint, diluted by a 179\,s exposure, 
	would be seen as having a visual magnitude $\sim 19$.
	
	Such flashes can be seen even at high declination, 
	and can be caused by satellites or space debris 
	from satellites launched into high orbits, 
	not necessarily on the equatorial plane, 
	e.g., Molniya and Tundra orbits. 
	The rate of glints at high declinations
	was measured by \cite{satellite_glints_EvryScope_Corbett_2020}
	that find a rate of $\approx 1000$ per hour per sky ($2.4\times 10^4$ per day), 
	for flashes close to the South celestial pole. 
	That survey was only sensitive to flashes with 
	peak magnitude brighter than $\sim 9$, 
	so that the rate of fainter glints, 
	similar in brightness to GN-z11-flash, 
	may be higher by an order of magnitude.\footnote{
		The rate in the equatorial region found by
		\cite{geosync_satellite_foreground_Nir_2020} 
		is about an order of magnitude larger than that found by
		\cite{satellite_glints_EvryScope_Corbett_2020}, 
		most likely due to the deeper limiting magnitude of the former. 
		This suggests that the rate of glints similar in brightness to GN-z11-flash 
		at high declinations could be 
		$\approx 10$ times larger than that given by 
		\cite{satellite_glints_EvryScope_Corbett_2020}. 
	}

	The rate of GRBs across the entire sky 
	is $\mathcal{O}(1)$ per day. 
	This rate is lower than the glint rate
	by four orders of magnitude. 
	If we assume GRBs occur only in galaxies, 
	and narrow the search region only
	to the surface area covered by high redshift galaxies
	(about 1\% of the sky\footnote{
		To estimate the angular area that could be associated 
		with a galaxy, we can assume that each galaxy takes 
		up an area $\sim 0.5$\,arcsec$^2$, including the size of the galaxy
		and the width of the seeing. 
		\cite{hubble_deep_field_galaxy_counts_Gardner_2000} estimate 
		the number of galaxies in the Hubble Deep Field 
		to be on the order of 500\,arcmin$^{-2}$, 
		so the total sky area covered by galaxies 
		is on the order of a few percent.
	})
	the rate of satellite flashes on top 
	of those galaxies would still be two orders of magnitude 
	higher than the rate of GRBs. 
	
	In addition, \cite{shock_breakout_z11_flash_Padmanabhan_2021} 
	suggested that GN-z11-flash is caused by a shock breakout event
	in a population III star in GN-z11. 
	However, this, as well as any physical model, requires the release 
	of about $4.5\times 10^{48}$\,erg, in $\sim 15$--35\,s, 
	in a narrow band between $\sim 1700$ and 2000\,\AA~(all numbers given in the rest frame).  
	
	%
	
	
	\section{Summary}
	
	A few conclusions based on these recent results are: 
	
	\begin{itemize}
		\item Glints can be seen at the position of GN-z11-flash, as high orbit satellites would be above Earth's shadow at that time and direction. 
		\item Short duration flashes from rotating objects can cause flashes shorter than 0.1\,s, that would appear point-like under $0.6''$ seeing conditions. 
		\item High orbit satellites do have high declination orbits. The rate of glints from such objects is at least $2.4\times 10^4$ per day per sky, 
		it is most likely higher by at least an order of magnitude, considering that fainter glints are more abundant. 
		\item The rate of such glints coincident with the area of distant galaxies on the sky is on the order of 200 per day per sky. 
	\end{itemize}
	
	Thus we suggest that given the current information, 
	a satellite origin can not be ruled out, 
	and the rate of such events is 
	higher than that of GRBs. 
	
	\pagebreak
	
	\bibliographystyle{aasjournal}
	\bibliography{refs}

\begin{thebibliography}{}
\expandafter\ifx\csname natexlab\endcsname\relax\def\natexlab#1{#1}\fi
\providecommand{\url}[1]{\href{#1}{#1}}
\providecommand{\dodoi}[1]{doi:~\href{http://doi.org/#1}{\nolinkurl{#1}}}
\providecommand{\doeprint}[1]{\href{http://ascl.net/#1}{\nolinkurl{http://ascl.net/#1}}}
\providecommand{\doarXiv}[1]{\href{https://arxiv.org/abs/#1}{\nolinkurl{https://arxiv.org/abs/#1}}}

\bibitem[{{Corbett} {et~al.}(2020){Corbett}, {Law}, {Vasquez Soto}, {Howard},
  {Glazier}, {Gonzalez}, {Ratzloff}, {Galliher}, {Fors}, \&
  {Quimby}}]{satellite_glints_EvryScope_Corbett_2020}
{Corbett}, H., {Law}, N.~M., {Vasquez Soto}, A., {et~al.} 2020, arXiv e-prints,
  arXiv:2011.02495.
\newblock \doarXiv{2011.02495}

\bibitem[{{Gardner} \&
  {Satyapal}(2000)}]{hubble_deep_field_galaxy_counts_Gardner_2000}
{Gardner}, J.~P., \& {Satyapal}, S. 2000, \aj, 119, 2589,
  \dodoi{10.1086/301368}

\bibitem[{{Jiang} {et~al.}(2020){Jiang}, {Wang}, {Zhang}, {Kashikawa}, {Ho},
  {Cai}, {Egami}, {Walth}, {Yang}, {Zhang}, \&
  {Zhao}}]{GRB_flash_high_redshift_galaxy_Jiang_2020}
{Jiang}, L., {Wang}, S., {Zhang}, B., {et~al.} 2020, Nature Astronomy,
  \dodoi{10.1038/s41550-020-01266-z}

\bibitem[{Jiang {et~al.}(2021)Jiang, Wang, Zhang, Kashikawa, Ho, Cai, Egami,
  Walth, Yang, Zhang, \& Zhao}]{refuting_more_probable_explanation_Jiang_2021}
Jiang, L., Wang, S., Zhang, B., {et~al.} 2021, "A more probable explanation" is
  still impossible to explain GN-z11-flash: in response to Steinhardt et al.
  (arXiv:2101.12738).
\newblock \doarXiv{2102.01239}

\bibitem[{{Kann} {et~al.}(2020){Kann}, {Blazek}, {de Ugarte Postigo}, \&
  {Th{\"o}ne}}]{GRB_flash_context_Kahn_2020}
{Kann}, D.~A., {Blazek}, M., {de Ugarte Postigo}, A., \& {Th{\"o}ne}, C.~C.
  2020, Research Notes of the American Astronomical Society, 4, 247,
  \dodoi{10.3847/2515-5172/abd4de}

\bibitem[{McLean {et~al.}(2008)McLean, Steidel, Matthews, Epps, \&
  Adkins}]{MOSFIRE_system_overview_McLean_2008}
McLean, I.~S., Steidel, C.~C., Matthews, K., Epps, H., \& Adkins, S.~M. 2008,
  in Ground-based and Airborne Instrumentation for Astronomy II, ed. I.~S.
  McLean \& M.~M. Casali, Vol. 7014, International Society for Optics and
  Photonics (SPIE), 1061 -- 1072, \dodoi{10.1117/12.788142}

\bibitem[{{Nir} {et~al.}(2020){Nir}, {Ofek}, {Ben-Ami}, {Segev}, {Polishook},
  \& {Manulis}}]{geosync_satellite_foreground_Nir_2020}
{Nir}, G., {Ofek}, E.~O., {Ben-Ami}, S., {et~al.} 2020, arXiv e-prints,
  arXiv:2011.03497.
\newblock \doarXiv{2011.03497}

\bibitem[{{Oesch} {et~al.}(2016){Oesch}, {Brammer}, {van Dokkum},
  {Illingworth}, {Bouwens}, {Labb{\'e}}, {Franx}, {Momcheva}, {Ashby}, {Fazio},
  {Gonzalez}, {Holden}, {Magee}, {Skelton}, {Smit}, {Spitler}, {Trenti}, \&
  {Willner}}]{high_redshift_galaxy_Oesch_2016}
{Oesch}, P.~A., {Brammer}, G., {van Dokkum}, P.~G., {et~al.} 2016, \apj, 819,
  129, \dodoi{10.3847/0004-637X/819/2/129}

\bibitem[{{Padmanabhan} \&
  {Loeb}(2021)}]{shock_breakout_z11_flash_Padmanabhan_2021}
{Padmanabhan}, H., \& {Loeb}, A. 2021, arXiv e-prints, arXiv:2101.12222.
\newblock \doarXiv{2101.12222}

\bibitem[{Steinhardt {et~al.}(2021)Steinhardt, Andersen, Brammer, Christensen,
  Fynbo, Milvang-Jensen, Oesch, \&
  Toft}]{more_probable_explanation_Steinhardt_2021}
Steinhardt, C.~L., Andersen, M.~I., Brammer, G.~B., {et~al.} 2021, A more
  probable explanation for a continuum flash in the direction of a redshift
  $\approx$ 11 galaxy.
\newblock \doarXiv{2101.12738}

\end{thebibliography}

\end{document}